\documentstyle[12pt]{article}
\begin{document}
\baselineskip=20pt

\title{Perturbations of Noise: The Origins of
Isothermal Flows}

\author{Piotr Garbaczewski\\
Institute of Theoretical Physics, University of Wroc{\l}aw, \\
pl. M. Borna 9,  PL-50 204 Wroc{\l}aw, Poland\\
 and \\
 Institute of Physics, Pedagogical University,\\
  pl. S{\l}owia\'{n}ski 6, PL-65 069 Zielona G\'{o}ra,
  Poland\thanks{Address from 1 Oct. 1998.
  Email: pgar@omega.im.wsp.zgora.pl}}
\maketitle
\centerline{PACS numbers:  02.50-r, 05.20+j, 03.65-w}
\begin{abstract}
We make a detailed analysis of both phenomenological and
analytic background   for the "Brownian recoil principle"
hypothesis (Phys. Rev. A 46, (1992), 4634).
A corresponding theory 
of the isothermal Brownian motion of particle ensembles
(Smoluchowski diffusion process approximation),
gives account  of the environmental
recoil effects due to locally induced tiny heat flows.
By means of local expectation values  we elevate the
individually negligible phenomena
to a non-negligible (accumulated) recoil effect on the
ensemble average.
  The main technical input is a consequent exploitation of
  the Hamilton-Jacobi equation as a  natural substitute  for the
local momentum conservation law. Together with the  continuity
 equation  (alternatively, Fokker-Planck), it forms a closed system
of partial differential equations which   uniquely determines
an  associated Markovian diffusion process.
The third Newton law in the mean is utilised to generate
diffusion-type processes which are either anomalous (enhanced), or
generically non-dispersive.

\end{abstract}
\vskip2cm

\section{Smoluchowski  diffusion processes, Stokes resistance
 and weakly out-of-equilibrium systems}

\subsection{Traditional phenomenology}

In random media that at are statistically at rest, motion  of
single tracers or dispersion of pollutants,
also in the presence of external conservative force fields, may be
consistently  described  in terms of Smoluchowski diffusion
processes, \cite{smol,nel}.  Their forward drifts are  interpreted
to give account of the mean velocity $\beta ^{-1}\vec{F}/m$
imparted by the external conservative force  to diffusing
particles,  an   outcome of the acceleration
$\vec{F}/m$ "felt" and accumulated on the relaxation time
scale $\beta ^{-1}$.    This time scale is regarded to be
much finer than the  one appropriate for the
coarse-grained   description. The latter,  $\triangle t \gg
\beta ^{-1}$, is still significantly  smaller than the
available  phenomenological (observational) resolution.

 Our basic intuitions are  rooted  in the
theory of a  Brownian  motion suitable for the description
of colloidal particles floating  in a liquid.
However an issue of  particular "causes of diffusion"  is not that
relevant, and  the proper  arena for the Brownian motion
can be not only a viscous  fluid, or  a dilute gas, but any
interacting many-particle system  and  even  less specific
random medium with a suitable microscopic  behaviour.
The problem  of "how to make a heat bath" needs to be addressed,
\cite{nel,cercignani,ford}.
In the present paper, we  take for granted the validity (even if
  diminished  to the status of an approximate theory or the
conceptual   playground)
of the standard Kramers and Smoluchowski diffusion process
scenarios, as the reference mathematical models of random
transport in the equilibrium bath.

If we consider  a   fluid in thermal equilibrium as the noise
 carrier, a kinetic theory viewpoint  amounts to visualizing the
 constituent molecules that collide
not only with each other but  also  with the tagged (colloidal)
particle, so \it enforcing \rm  its  observed erratic  motion.
Clearly,  random molecular collisions  both
initiate and maintain an incessant irregular motion of the tagged
particle.
 Once the particle is in motion,
 we need to account for an  additional  statistical
effect  of molecular impacts  on the  actually \it moving \rm
 particle. It is  phenomenologically  encoded in  the
Stokes resistance, coming from averaging over a molecular "rain"
along some portion of the trajectory, which is
 proportional to the velocity of the particle.
In the phase-space description,  the particle velocity
$\vec{V}(t)=\vec{v}$ is a random quantity.
The damping effect is  \it not \rm  random.
It stands for a statistically
accumulated, passive response of the medium. Indeed (see e.g.
Section 2.1 in below),
$-{\vec{v}\over {\beta ^{-1}}}$
 locally averages the genuine (ignored fine-detailed) random
 dynamics  on the relaxation time scale $\beta ^{-1}$.

For a bath in  equilibrium, the internal relationship
between the above random and systematic (due to friction)
effects  of a generic noise on the Brownian particle motion
is the subject of fluctuation-dissipation theorems.
This   extends to situations when the external driving forces
additionally modify the particle dynamics, but do not modify the
statistics of noise (!).  This feature is generic to
Smoluchowski diffusions.
(Conventionally, the  fluctuation-dissipation theorems are
interpreted, \cite{kubo},
to set a general relationship between  the response of
diffusig particles in an equilibrium bath
to an external force,  and the internal fluctuation   of the
bath in the absence of this  disturbance.)

By means of  Einstein's fluctuation-dissipation theorem we
are  given  the diffusion coefficient  $D={{kT}\over {m\beta }}$.
It  characterises an intensity of the spatial (Wiener) noise
 in terms of the friction parameter $\beta $,
 with $k$ the  Boltzmann constant, $T$ the equilibrium temperature
 of the bath, and $m$ mass of a diffusing particle, \cite{smol}.
A  formal  exploitation of  the Stokes formula
(derivable on the basis of pure kinetic arguments from
the Boltzmann theory, \cite{dorfman})  transfers to the Brownian
realm  a concept of  a  frictional deceleration, originally
suited for a macroscopic  spherical
particle of radius $a$ and mass $m$ moving in a  fluid with
viscosity coefficient  $\eta $.  In the new context, it is
the   \it mean \rm property of motion and \it not \rm a
particular single  Brownian particle-in-motion  attribute.
Nonetheless, that allows to establish   the value of the
friction constant   $\beta = {{6\pi \eta a}/m}$,  the  result
 amenable to positive experimental verifications in the classic
 studies of the Brownian motion, \cite{smol,perrin}.

\subsection{Meaning of stochastic models}

An  observable Brownian motion of  colloidal particles in a fluid,
\cite{smol,perrin}, when interpreted in terms of random
processes, involves a number of mathematical subtleties, like e.g.
 an inherent  nondifferentiability of
 sample paths in  velocity space that reappears on the
 spatial arena  of Smoluchowski processes as well.

Clearly, the  phase-space stochastic process is  rather
crude approximation of reality, if compared with the kinetic
theory
reasoning based on the  explicit input of particle collisions
 to the Boltzmann equation  and the related  kinetic theory of
hydrodynamic flows, \cite{dorfman}.  The Boltzmann equation
can  be interpreted in terms of a jump Markov process simulating
collision events, and the diffusion-type  Karmers equation may
arise only  in a suitable scaling limit, \cite{tanaka}.  This
limit can be  justified  in case of grazing collisions,  or as a
way to include the effect of long range forces  by treating
their influence in a statistical way: they should generically
produce a sequence of small and almost random changes in the
tracer particle velocity, \cite{cercignani}.

The Smoluchowski \it approximation \rm takes  us  further away
 from the kinetic theory intuitions by  projecting  the
phase-space theory of random motions into its  configuration
space image, \cite{wil},  which is a spatial Markovian diffusion
process.

To quantify  the above picture  one usually departs from the
phase-space (Langevin and Kramers) description  of fluctuation
phenomena.
Let us consider an It\^{o} equation (in  its symbolic differential
version) for infinitesimal increments of the velocity random
variable, exhibiting the systematic frictional resistance:
$${d\vec{V}(t)=-\beta \vec{V}(t)dt + \beta \sqrt{2D} d\vec{W}(t)}
\eqno (1)$$
where $\vec{W}(t)$ stands  for the normalised Wiener process.
One can  easily infer, \cite{horst},  the corresponding second
Kolmogorov (Fokker-Planck) equation
$${\partial _t p(\vec{v}_0,\vec{v},t) = \beta D \triangle _{\vec{v}}
+ \beta \nabla _{\vec{v}}\cdot [\vec{v}\,  p(\vec{v}_0,\vec{v},t)]}
\eqno (2)$$
for the transition probability density of the time homogeneous
process in the velocity space alone, \cite{horst,nel}.

In view of:
$${p(\vec{v}_0,\vec{v},t) =
({m\over {2\pi kT(1- e^{-2\beta t)}}})^{3/2}
exp \{ {m\over {2kT}} {{(\vec{v} - \vec{v}_0e^{-\beta t})^2}
\over {1-e^{-\beta t}}}\} }\eqno (3)$$
the time interval $\beta ^{-1}$  effectively  accounts for
an approach of the transition  density
 to the equilibrium  Maxwell distribution.

Let us consider an instantaneous velocity  $\vec{V}(t)=\vec{v}$,
 that  has been achieved in the course of the  random evolution (1)
 beginning from a  certain $\vec{V}(0)=\vec{v}_0$.
We can evaluate a conditional expectation value (local mean with
 respect to the law of random displacements (3)) over all randomly
 accessible velocities   $\vec{V}(t+\triangle t)=\vec{v}'$
at a  time  $t+\triangle t,\,  \triangle t>0 $. It determines
the  forward drift of the process:
$${\vec{b}(\vec{v},t)= lim_{\triangle t \downarrow 0}
[\int \vec{v}'\,
p(\vec{v},\vec{v}',\triangle t)\, d^3v' - \vec{v}] = -\beta \vec{v}}
\eqno (4)$$
and  thus, \cite{nel,vigier,blanch},  provides us with an information
about the \it mean tendency \rm
of the dynamics  on  small  (but not too small if compared to
 $\beta ^{-1}$) time scales.
Evidently,  that    mean  tendency in case of Eq. (1)  is to
decelerate the velocity $\vec{v}$ at the Stokes rate
${\vec{v}\over {\beta ^{-1}}}$.

Sample paths of the Wiener process are  nondifferentiable and play
the r\^{o}le of  (velocity space) idealizations of  "true" trajectories.
We need a conditional averaging over all  paths
emanating  from $\vec{v}= \vec{V}(t)$  to bypass, cf. Eq.(4),
the nondifferentiability problem and then introduce the local
 deceleration rate  ($ \vec{v} \rightarrow \vec{v} -
\beta \vec{v} \triangle t$), \cite{nel}.
Clearly, the friction term and thus the  deceleration concept,
in the framework set by Eq. (1), arise as   statistical,
\it local mean value\rm ,  quantities.
All that must
be sharply contrasted with the standard hydrodynamical meaning of
the Stokes resistance formula, which refers to  a genuine
(the reduced Navier-Stokes equation is involved) fluid
velocity field  around a \it single  \rm uniformly  propagating
particle.
It is the perturbation by the moving macroscopic
body that gives rise to the
 force, with which the viscous fluid acts upon the concrete
 particle-like object.

Quite apart from  the time-scale
($\triangle t $ versus $\beta ^{-1}$) issue, an individual
Brownian  particle neither moves uniformly, nor \it ever \rm
permanently relaxes to the  state of rest.
Actually, for a single Brownian particle, the respective
velocity gain/loss (acceleration/deceleration rate) on
the $\beta ^{-1}$ time scale \it is a random quantity. \rm

 On the contrary, the standard
interpretation of the Stokes resistance   for a large   particle
in a viscous fluid, makes inavoidable  an  ultimate stopping of the
particle,  unless   an external force would balance    the
damping and so maintain the (uniform)  particle motion at
a certain terminal velocity. This  latter concept forms the
 basis of the Einstein  fluctuation-dissipation formula in case of
 the gravitationally induced sedimentation phenomenon,
 cf. \cite{smol}, albeit in case of the Brownian motion  only  local
 mean velocity fields  (hence, ensemble averages) can be employed
 for that purpose.
 The picture of "a Brownian particle moving at its terminal velocity",
 \cite{streater}, is certainly inappropriate.

We can supplement  Eq. (1) by the spatial increment definition:
$d\vec{X}(t) = \vec{V}(t)dt$, extending (1) to the phase-space
process whose Smoluchowski  projection in the large damping
regime reduces  to the pure spatial Wiener noise
$d\vec{X}(t)=\sqrt{2D}d\vec{W}(t)$.
In this  case, the  generic   spatial
 scale (wandering distance over which  the  dissipation
process \it in the mean \rm is completed)   is  set by
$(D\beta ^{-1})^{1/2}$.   The Smoluchowski approximation
amounts to ignoring  the fine details of the dynamics
on  the $\beta ^{-1}$ and $(D\beta ^{-1})^{1/2}$   scales.
One disregards  all possible remnants of the phase
space process that extend  beyond those  scales (the damping
typically
induces $exp(-\beta t)$ factors in all relevant formulas, hence
a  certain  amount  of "memory" must be
eliminated to yield a  Markov process), \cite{smol}.

That  particular  \it disregard/ignore \rm issue is worth
emphasizing
in the context of typical approximate (analytic or numerical)
manipulations   with the Brownian motion.
A  strong solution of Eq. (1)   (we take
$\vec{V}(0)= \vec{v}_0$ as the initial condition) has the form :
$${\vec{V}(t)=\vec{v}_0 exp(-\beta t) + \int_0^t exp[-\beta (t-s)]
\sqrt{2D} d\vec{W}(s)\enspace .}\eqno(5)$$

This  expression, albeit   looking  physically
unrealistic,  is a fairly accurate  approximation of a
phenomenological  molecular collision scenario for the Brownian
motion. Namely, let us introduce a finite Riemann sum
approximation of the integral in Eq. (5):
$${\int_0^t exp[-(t-s)] \sqrt{2D} d\vec{W}(s)
\approx exp(-\beta t) \sum_n
exp(\beta n\triangle t) \triangle \vec{V}_n}\eqno (6)$$
where  $\triangle \vec{V}_n= \sqrt{2D} \triangle
\vec{W}(n\triangle t)$ stands for the
n-th consecutive velocity  increment, e.g. an effect of all "random
acceleration" events taking place in the time interval
$(n\triangle t,(n+1)\triangle t)$.
If we resort to a   molecular collision mechanism (in
a dilute gas for example),
the velocity increments due to collisions of the Brownian  particle
with molecules  of the bath can be viewed as statistically
independent, and occur at an  approximate rate  $10^{-21}$ per second.
The collisions may be interpreted to occur consecutively one after
another, but  multiple  collision events are  allowed as well.
If we regard   the time coarse-graining in Eq. (6) to refer to
a typical relaxation time $\beta ^{-1}\approx 10^{-8}s$,
(notice that the respective coarse-graining appropriate for the
Smoluchowski approximation, would involve
$\triangle t\gg \beta ^{-1}$), it is obvious that
each n-th velocity increment can be interpreted as
a sum  of an enormous number  of  independent identically
distributed random variables  (minute acceleration/deceleration,
 e.g. collision   events).  Let us stress that we exploit here
 a defining property of the  Wiener process; the velocity
 random variable (5)  of the Ornstein-Uhlenbeck process (1)-(3)
 is known \it not \rm to have independent  random increments,
 \cite{horst}.

The accumulated fluctuation irregularities,
on the $\beta ^{-1}$ time scale,  are  the primary
reason of the erratic behaviour of the Brownian motion in the
velocity space.
Thus, an  integration with respect to the normalised Wiener process
$\vec{W}(t)$ in Eq. (5),  quite satisfactorily (in a suitable
scaling limit)  models a \it cumulative \rm   outcome
of phenomenologically motivated  impulses (kicks exerted by
the noise carrier) on the Brownian
particle.

Let us emphasize at this point, \cite{nel},  that it is not correct
to think  that conspicuous jiggles
in the Brownian trajectory   are due to \it single \rm kicks.
(In  mathematical  terms the
situation is even worse, since the Wiener process has an unbounded
variation on aritrary time intervals.)
The realistic Brownian motion is unbelievably gentle, specifically
if we refer to  a heavy particle immersed in a gas of much
lighter ones.
Each collision (kick) has an entirely negligible effect  on the
velocity  of the Brownian particle.
It is only fluctuations in the accumulation of an enormous
number of very slight  changes in
the particle velocity which give the trajectory its irregular
appearence, both in velocity and position pictorial
representations of motion.   Conequently, equation (5) makes sense
as a useful approximation of physical phenomena,
if the coarse-grained   time scale $\triangle t$ in (6)
 (which is far below an observational one)
is much larger than $\beta ^{-1}$, but still small enough for
a  sufficiently fine coarse-graining  (6) of (5).

The previously mentioned  space and time scales  justify the
utility of the Smoluchowski description of conservative force
effects upon a spatially diffusing particle,  \cite{nel}:
$${d\vec{X}(t)= {\vec{F}\over {m\beta }}dt +
\sqrt{2D}d\vec{W}(t)\enspace .}\eqno (7)$$

The Smoluchowski forward drift can be traced back to
a presumed selective action of the force on the Brownian particle
that  has a negligible  effect on the thermal bath.
Indeed, we take for granted that there  is  no physically  relevant
mean  (induced)  flow in the bath proper,
unless the isothermality assumption is abandoned,
\cite{streater,mazur1},
or  other  hitherto  disregarded effects on the $\beta ^{-1}$ scale
 (like those due to the action-reaction
 principle, \cite{vigier})  are incorporated into the formalism.

Brownian particles appear to be  drifting  \it on the local
average \rm relative to the bath,   with a uniform (but in
the mean !) velocity ${\vec{F}\over {m\beta }}$. Clearly, a
 repeated series of  observations, at regularly spaced
 time intervals,   of a \it single \rm tagged  Brownian particle
 would not in general  reveal  any specific motion tendency.
The graphical  picture of motion   would be as irregular as ever
("no purpose" sample
  paths of the Wiener process, like  in the absence of any force).
The respective  coarse-grained
approximation of  the trajectory does  correspond  to an
experimental resolution,  which  is incomparably rougher than
the previous  time scales.

Only a  numerical  simulation of the  statistical ensemble of
sample paths with  a  controlled probability distribution
(frequency, in fact) of initial conditions, or a realistic
monitoring  of the  Brownian motion-induced dispersion of
a low density cloud of  dust particles suspended in a liquid,
would lend a definite meaning to the "motion tendency" concept
and to the related mean Brownian flows.
Realistic diffusion
processes (and diffusive transport) are observed in case of
non-uniform concentrations of
colloidal   particles but they are regarded  as a result
of random migration of individual particles which is
actually observable (under a microscope) and interpreted  as
the Brownian motion.

In the context of a sedimentation phenomenon,  a  sufficiently
long overall observation time  of a single tagged   particle
travelling along its  erratic path (hours or days in a real,
or computer simulation experiment),  would presumably reveal
that a  particle more frequently visits certain   spatial areas,
in accordance with the barometric formula (ergodic features of
motion).
Early experiments on this issue, \cite{smol,perrin}, pertained to a
cloud of suspended particles executing  an extremely slow
(practically adiabatic) diffusion process.  In some cases,
the spatial displacement  of the size $0.2 \cdot 10^{-6} m$
tracer particles  has been measured in  $30 s$ time intervals,
\cite{perrin,smol}, to yield the observationally relevant
outcomes. Very recent  observations of  an
individual Brownian particle motion refer to
$1-2.5\cdot 10^{-6}m$ tracers and the ${1\over {60}} s$
time resolution.

\subsection{Problems with thermal equilibrium}

A spectacular solution  of the sedimentation problem due to
Smoluchowski, refers to the isothermal dynamics  (1) which is
constrained to the positive vertical semiaxis:
$dZ(t)= -\beta c dt + \beta \sqrt{2D} dW(t)$,
 with a  reflecting boundary at the spatial point $0$.
The transition probability density of this one-dimensional
 process reads, \cite{smol,horst}:
$${p(z,z_0,t)= {1\over {2(\pi Dt)^{1/2}}}\{ exp[-(z-z_0)^2/4Dt] +
exp[-(z+z_0)^2/4Dt]\}  +}$$
$${{c\over {D\sqrt{\pi }}}exp(-cz/D)\,    \int_{{z+z_0-ct}
\over {2(Dt)^{1/2}}}^{\infty } exp(-x^2)dx}\eqno (8)$$
and shows that a particle starting its motion from any positive
$z_0$ near $0$, may wander along the positive semiaxis indefinitely.
In particular, it can be transported against the  gravitational
force to an arbitrary height.  Depending
on the actual $z_0$ the Brownian particle may  have a higher
probability to ascend than to descend.

The corresponding \it mean work \rm
has been evaluated to be  $kT$ per particle, cf. \cite{smol,horst}
for a related discussion of the entropy decrease issue.  Certainly,
an ability of the medium to perform work (in the  mean) i. e. to
give a kinetic  energy to  the Brownian particle (on the ensemble
average  again), is not special to the sedimentation problem.
 It  appears to be a universal feature of  the thermal bath even in
 the absence of any external forces.

In the  equilibrium situation ($t\rightarrow \infty \Rightarrow
p(z,z_0,t)\rightarrow \rho (z)$), we would
arrive at the familiar balance condition:  the mean tendency of
motion  (forward drift) due to gravitational acceleration
must be  exactly balanced by the oppositely directed motion tendency
of the diffusive (osmotic pressure, \cite{smol,nel})  origin,
valid for non-uniform concentrations of a contaminant in a solvent.
The latter tendency involves
 sending particles away from the areas of higher probability of
 their presence
(concentration, if a low density pollutant is considered)
in accordance with the Fick formula for the diffusion current
$- c= -  D {{\nabla \rho (z)} \over {\rho (z)}}$.
The barometric formula does  follow.

It is interesting to spend a while on some
tagged (single) particle features in the nonequilibrium--but
isothermal--regime admitted by (8) for not too large times,
when the Einstein (mean) balance condition is yet invalid.
Clearly, to have defined an analogue of the  Fick diffusion
flow, a probability density  of initial data $z_0$
must be chosen. In a computer simulation we would have under
 control a bunch of    relevant sample paths (all consecutively
 executed on a fixed duration time interval $[0,t]$)
 and the related probability density evolution along the bunch.
 Depending
 on the   initial data distribution, for a  time period  the
 osmotic  drift would dominate  the gravitationally
 induced drift.  To this end,  work   (in the mean)  must have been
  done on  Brownian  particles \it at the  expense of the bath. \rm
 Local conversion of work into heat seems to be inavoidable, and
 local heat flows are inavoidable as well, cf. \cite{streater,perrin}.
That creates a number of problems to which no attention is normally
paid in the literature.

Smoluchowski diffusions are conventionally
regarded as isothermal processes (possible heat flows
are ignored by various reasons). If we however admit that the
emerging tiny heat flows may  have an affect on the particle
transport due to the Brownian motion, a suitable description of the
thermal inhomogeneities and their   effects on the dispersion of
Brownian particles must be invented.

For example,
in case of thermally  inhomogeneous gases it is well known
that a dust-free region appears about a hot body, showing
that a temperature gradient has  an effect
on the motion of  dust particles in a gas, \cite{mason}.
Tracer particles are transported away from the hot areas
to  the cooler ones (we may interpret that as a repulsion by
the  heated domain). On the reverse, they appear to be
 attracted by the cooler areas while  escaping  from
 the hot domains.  Particles float down the temperature
 gradients.

\vskip0.2cm
{\bf Remark 1:}
Let us mention an approach to quite similiar problem,
\cite{streater} which  was  originally formulated  for a
cloud  of contaminants in a liquid,   under the following
phenomenological assumption:  "a gas of  Brownian  particles falling
in gravity should leave a trail of warm fluid in its wake, since its
potential energy is being converted into heat".
Obviously,  if the  particles would move against  gravitational force,
then the temperature of the medium  should locally drop down.
Those features, if we are to keep track of the local heating and
cooling (as opposed to the isothermal Einstein or Smoluchowski
diffusive dynamics) were interpreted as a source of the space-time
dependence of temperature. The Fokker-Planck equation must then
be supplemented by  an evolution equation for the temperature field,
(a clear-cut kinetic theory reasoning can be read out in this
strategy), so that the coupled nonlinear system  would take the
 form of a "consistent  thermodynamical system"  i. e. the one
 manifestly respecting  the
 first and second  laws of thermodynamics, see e. g. Ref.
 \cite{mazur1}.
 Here, the heat flows are assumed to be neither slow nor fast
 enough to be effectively disregarded. \\

\vskip0.2cm
{\bf Remark 2:}
An issue of suitable slow and fast process time scales is crucial
in our  discussion.
One should be aware that  local temperature perturbations  of the
bath
may be safely neglected  when  the dissipation (fast process)
time scale and the diffusion  (observational for tagged particles)
time scales are
generically incompatible, like   e.g. in case of a rapid
dissipation set
against a slow diffusion process.   Then,  the usual  isothermal
diffusion process  follows and the standard Brownian motion
paradigm is left intact.
\vskip0.2cm

In the discussion of the Carnot principle, in reference to the free
Brownian motion and to the sedimentation problem, \cite{smol,perrin},
 Brownian particle fluctuations are regarded to occur due to
 causes that  are intrinsic to the random medium.
If we think about minute acceleration/deceleration events  that
modify (say,  at a rate of $10^{21}$ times  per second)
velocities of  realistic    particles,  the microscopic
energy-momentum  conservation laws need to be respected in
each separate collision event.  In contrast to derivations based
on the Boltzmann colllision scenario, this feature is completely
\it  alien \rm to the Brownian motion theory,
cf. \cite{vigier,blanch}.
(This happens quite apart from the elusive
power of the flutuation-dissipation mechanism, \cite{kubo}:
"The friction, or more generally the resistance of  a given system,
represents the method by which the external work  is dissipated
into microscopic  thermal energy. The reverse process is the
generation  of random force as the result of thermal fluctuation".)

Let us point out that the phenomenology allowing to regard  Eq. (1)
as a satisfactory model
for Brownian particle velocity fluctuations, blurs possible
advantages of   the closely related  argument appearing in
the diffusion modeling of
the Raileyigh piston, \cite{miller}: " Collisions between the
large particle (piston)  and the
bath particles are elastic. After each collision the bath particles
are given a new distribution, which evolves until the next collision".

To avoid an apparent contradiction with   the law of
energy   conservation, there seems tempting to require that
each minute acceleration of a Brownian particle   is accompanied
by a minute \it cooling \rm of the medium in its immediate
neighbourhood.
Correspondingly, any deceleration event should
induce a local \it heating  \rm of the immediate neighbourhood
of a particle, see e.g. \cite{perrin}.     Since (cf. Eq. (1)), we
always disregard the fine details of about $10^{13}$ collision
impacts on the Brownian particle on a typical relaxation
time scale of $10^{-8}s$, there is  definitely enough  room to
allow for local  statistical measures of heating and cooling.

\vskip0.2cm
{\bf Remark 3:}
The heating, cooling
and  temperature notions are quantitative mean measures of
the degree  of agitation of the noise carrier.
If the random medium is interpreted
on the molecular level to be composed of  light particles,
these measures  can be  related with the  mean square
deviation of the (bath molecule) velocity random variable,
or an average of the squared velocity if its mean value  vanishes.
  Those quanties are purely statistical characteristics of the bath
  and \it not \rm of the Brownian particle immersed in it.
Only under restrictive thermal equilibrium conditions, the notion
of temperature appropriate to the bath,  can be elevated to the status
of a  measure of a thermal agitation for tracer particles.

\subsection{Goals}

In the Kramers approach to the phase-space dynamics, the
stochastic properties  of the medium  were  considered to be
independent from  random
phase-space data of the  Brownian particle. The
statistics of  noise (e. g. the thermal
equilibrium features of the bath) must have remained
 unperturbed by the very presence of the particle
and its phase-space  fluctuations,
albeit those   are \it enforced  \rm  by the   intrinsic
randomness of the bath.   One assumes that there is no
relevant \it dynamical \rm  response of the bath  to the very
presence of  the Brownian particle and its induced dynamics.
We recall that the fluctuation-dissipation theorems merely
account  for a \it passive \rm response,
in terms of the  statistically implemented  Stokes
resistance of the bath to the particle-in-motion.

On the other hand, the above  local heating and  cooling notions
refer to the    dynamical response of the bath to the
Brownian particle which is immersed in it.
We need to account for the
out-of-equilibrium properties of the bath in the presence
of a single particle,  whose motion  is solely enforced
by the bath.  Thus
very weak (although possibly fast) heat flows  should
accompany an individual Brownian particle motion.
That conforms with an obvious
intuition that  no physical system is ever in thermodynamical
equilibrium, and normally it is necessary to idealize the
situation by regarding "fast processes" to be completed,
while the "slow" ones are  still running.

     In such a weakly non-equilibrium system
 with small heat flows, we may  expect that the
 standard equilibrium temperature notion is replaced
 by an \it effective \rm
 temperature notion (and an effective thermal equilibrium),
  which depends     on the chosen fast-versus-slow-process
  time scales and the ensemble averaging.
Repetitions of a single particle experiment in  the same
thermal bath, should now be replaced by
repetitions  of the same experiment with different realizations
of the out-of-equilibrium heat bath, cf. \cite{havlin} for similiar
concepts in the context of randomly disordered media.

In the above sense only  an effective isothermal regime
may be maintained, since the compensating heat flows allow to
regain the equilibrium temperature \it almost \rm instantaneously.
After  averaging over the tracer particle ensemble and the
corresponding
(weakly out-of-equilibrium)  sample paths,  we  should be able
to capture possible statistically relevant effects due to temperature
inhomogoneities (small deviations from thermal equilibrium conditions)
and  the resultant  effective flows  \it in a  bath,  \rm
that are possibly induced
(via back reaction on the time scale probably larger than
$\beta ^{-1}$ but still much below the Smoluchowski
approximation time scale $\triangle t$)  by
propagating Brownian particles. All that is to happen well
beyond the limits    of an available, cf. \cite{perrin},
 observational resolution.

We have thus set a phenomenology of a specific
\it dynamical \rm response of the  random medium  to the
Brownian motion    of a \it single \rm tagged particle,
whose sole outcome  are the minute deviations from the thermal
equilibrium of the fluctuating medium itself and the resultant
heat flows (needed to restore the equilibrium).
This   is independent from   the
traditional frictional resistance argument, directly referring
to the thermal equilibrium  conditions and the $\beta ^{-1}$
time scale.

We deal here with a  generic  \it feedback mechanism \rm scenario.
The Brownian particle  propagates "at the expense" of the   bath,
which however remains   "close" to its thermal equilibrium.
The bath in turn actively \it reacts back \rm to what is being
happenning to the particle  in the course of its propagation.
The  instantaneous  local deviations
from the state of equilibrium   ("perturbation of noise")
along the trajectory  surely  have  an  effect on each  subsequent
stage of   the particle propagation.
 Even if residual for an individual   Brownian
particle and its  sample path, the feedback effect is expected
to accumulate  statistically  (on the ensemble average)
 to a  sizable quantity.
  Under the name of the "Brownian recoil principle" we have made a
  preliminary study of such random dynamics in Refs.
  \cite{vigier,blanch}.

\vskip0.2cm
{\bf Remark 4:}
In connection with the previous Remark 1, let us mention that
a spatial diffusion (Smoluchowski) approximation of the
phase-space process, allows
to reduce the number of independent local conservation laws
(cf. \cite{mazur,zambrini,geilikman}) to two only.
Therefore the Fokker-Planck (or continuity) equation can
always be supplemented  by another (independent) partial
differential equation to form a closed system.
Non-isothermal flow description needs
to accomodate the variations of temperature of the bath,
(cf. \cite{streater,mazur1}), while we investigate the limits of
validity of the isothermal  scenario.
That amounts to inequivalent choices of the supplementary equation.
We emphasize  a single tagged particle-in a bath  description
in a repeatable experimentation sequence, under
basically  the same (or very similiar) physical conditions.
In such situation, a  stochastic process executed by a single
particle is practically isothermal
("the Brownian motion  is unbelievably gentle", \cite{nel}).
Each sample path  of the Brownian particle (and the related
cooling/heating phenomena induced  along the path)
is a random  quantity. The state of the medium, if giving account
of its \it dynamical response \rm (i. e. deviation from thermal
equilibrium) to the Brownian propagation, is a random
quantity as well. It is the local mean (related to the ensemble
average over various realisations of out-of-equilibrium
 conditions for the bath)   that may properly quantify
 this picture.

\section{Local conservation laws for the Brownian motion in the
Smoluchowski approximation: diffusion
currents and driving flows}

Before, we have identified   the forward drift
${\vec{F}\over {m\beta }},\,  \vec{F}= - \vec{\nabla }V$ as a
quantitative measure of
 a statistical  (local) tendency of the Brownian motion, obtained
 through averaging over  an  ensemble of sample paths.  If we assign
 a probability density $\rho _0(\vec{x})$ with which the
 initial data  $\vec{x}_0=\vec{X}(0)$ for Eq. (7) are
distributed (weak solutions of (7) enter the scene), then
the emergent Fick law  would  reveal a
statistical tendency of particles to flow away from
higher probability  residence areas.     This
feature is encoded in the corresponding Fokker-Planck equation
(equivalently, a continuity  equation):
$${\partial _t \rho  = - \vec{\nabla }\cdot (\vec{v}\rho )=
- \vec{\nabla }\cdot [({\vec{F}\over {m\beta }}  - D
{{\vec{\nabla }\rho }\over {\rho }}) \rho ]}\eqno (9)$$
where a diffusion current velocity  is
$\vec{v}(\vec{x},t) = \vec {b}(\vec{x},t) - D{{\vec{\nabla }\rho
(\vec{x},t)}\over {\rho (\vec{x},t)}}$
while  the forward drift reads $\vec{b}(\vec{x},t) =
{\vec{F}\over {m\beta }}$, cf. Eq. (7).
Clearly, the local diffusion current (a local flow that might
be experimentally observed for  a cloud of suspended particles
in a liquid)
$\vec{j}=\vec{v} \rho $ is
nonzero  in the nonequilibrium  situation and
a non-negligible  matter transport occurs as a
consequence of  the Brownian motion, on the ensemble average.

It is interesting to notice that the local velocity field
 $\vec{v}(\vec{x},t)$
  obeys the natural (local)  conservation law, which we
quite intentionally pattern after the  moment identities
(hierarchy of conservation laws) valid for the Boltzmann and Kramers
equations, \cite{wil,mazur}.
 The pertinent momentum conservation law  directly originates from
the rules of the It\^{o} calculus for Markovian diffusion processes,
\cite{nel}, and from the first moment equation in the
diffusion approximation  (!) of the Kramers theory, \cite{wil,mazur}:
$${\partial _t\vec{v} + (\vec{v} \cdot \vec{\nabla }) \vec{v} =
\vec{\nabla }(\Omega - Q)\enspace .}\eqno (10)$$

An effective  potential function $\Omega (\vec{x})$
can be expressed in terms of the  forward drift
$\vec{b}(\vec{x}) = {\vec{F}(\vec{x})
\over {m\beta }}$ as follows:
$${\Omega = {{\vec{F}^2} \over {2m^2\beta ^2}} + {D\over {m\beta }}
\vec{\nabla } \cdot \vec{F} \enspace .}\eqno (11)$$

Let us emphasize that it is the diffusion (Smoluchowski)
approximation, \cite{wil,mazur}, which makes
the right-hand-side of Eq. (10) substantially  different from the
usual moment equations appropriate for  the Brownian motion,
\cite{mazur}.  In particular, the force $\vec{F}$ presumed to act
upon an individual particle, does not give rise in Eq. (10)
to the  expression
$-{1\over m}\vec{\nabla }V$  which might be expected on the basis of
kinetic theory intuitions and moment identities directly derivable
from the Karmers equation, but to  the term
$+\vec{\nabla }\Omega $, cf. Eq. (11).

Moreover, instead of the standard pressure term,
 there appears a contribution from a
  probability density  $\rho $-dependent potential
 $Q(\vec{x},t)$. It is  is given in terms of the so-called osmotic
velocity field  $\vec{u}(\vec{x},t)$, (cf. \cite{nel}):
$${Q(\vec{x},t) = {1\over 2} \vec{u}^2 + D\vec{\nabla }
\cdot \vec{u}}\eqno (12)$$
$$\vec{u}(\vec{x},t) = D\vec{\nabla } \, ln \rho (\vec{x},t)$$
and  is  generic to a local momentum conservation
law  respected by   isothermal Markovian diffusion processes, cf.
 \cite{nel,vigier,blanch,zambrini}.
 Notice that in  case of the free Brownian motion (admitted, if
 we set $\Omega = 0$), we would have
 $\vec{v}(\vec{x},t) = - \vec{u}(\vec{x},t)$ for all times.

 An equivalent form of the potential (12):
 $Q= 2D^2{{\triangle \rho ^{1/2}} \over {\rho ^{1/2}}}$
 induces  rather obvious quantum mechanical associations (the de
 Broglie-Bohm "quantum potential" with the opposite sign, modulo
 an adjustment of constants), \cite{wilhelm}.  In the context of
 the Brownian motion, this "quantum potential" has been deduced in
 earlier  investigations of local conservation laws,
  \cite{geilikman}.
 \vskip0.2cm
 {\bf Remark 5:}
 Let us notice that by demanding $Q=\Omega $ identically for
 all $\vec{x}, t$, we would reduce Eq. (10) to
 $\partial _t\vec{v} +
 (\vec{v} \cdot \vec{\nabla }) \vec{v} = 0$. Despite of its
 classical-looking Riemann equation form, this conservation law
 still refers to a diffusion process. Namely, in view of (11),
 we must  identify forward drifts  with osmotic velocity fields
 and $D\vec{\nabla }ln\, \rho (\vec{x}) =
 {\vec{F}\over {m\beta }}$ holds true. The related diffusion
 process is stationary and preserves the  probability measure
 ($\rho $ is now time-independent). Cf. \cite{streit}
 for more general considerations  on that  issue.
\vskip0.2cm

As repeatedly stated before, Smoluchowski drifts refer to  mean
motions relative to the bath at rest, and there is no place for
any flows intrinsic to the random medium in this formalism.
 On the other hand, it is
of fundamental importance to understand how   genuine
flows  in a random medium  may be generated and what would be their
 effect on dispersion, \cite{blanch}.
(Solutions of the incompressible  Navier-Stokes equation may serve as
a common-sense model of the flow in a bath, and  the diffusion
enhancement is known to  be  related to various turbulent motion
scenarios.)

To  analyze  random
perturbations that are either superimposed upon or are intrinsic to a
driving deterministic  motion,
a configuration space equation
${\dot{\vec{x}}=\vec{w}(\vec{x},t)}$
 is normally invoked,  which  is next replaced by
a formal infinitesimal representation  of   an It\^{o} diffusion 
process
$${d\vec{X}(t)= \vec{b}(\vec{X}(t),t)dt + \sqrt{2D} d\vec{W}(t)}
\eqno (13)$$
patterned after Eq. (7). The tacit  assumption  (basically wrong,
\cite{muratore}) is
that $\vec{b}$ does not substantially differ from $\vec{w}$.

It is useful  to exploit
a standard phase-space argument that is valid, under isothermal
conditions,  for a Markovian diffusion process
taking place in (or relative to) a  flow  $\vec{w}(\vec{x},t)$
with as yet
unspecified dynamics nor concrete physical origin.
We account for an explicit force (here, acceleration
$\vec{K}= \vec{F}/m$)
exerted upon diffusing particles, while not  directly
affecting the driving flow itself.
Namely, \cite{horst,nel}, let us  set for infinitesimal increments of
phase-space random variables:
$$d\vec{X}(t)= \vec{V}(t) dt $$
$${d\vec{V}(t)= \beta [\vec{w}(\vec{x},t) - \vec{V}(t)] dt +
\vec{K}(\vec{x})dt  + \beta \sqrt{2D} d\vec{W}(t)\enspace .}
\eqno (14)$$

Following the leading idea of the Smoluchowski approximation, we
assume
 that $\beta $ is large, and consider the process on  time scales
 significantly exceeding $\beta ^{-1}$ (that is normally achieved
 by taking  $\beta $ to be very large, cf. the infinite friction
 limit procedure).
 Then, an appropriate
  choice of
 the velocity field $\vec{w}(\vec{x},t)$ may
 in principle guarantee, \cite{nel},  the convergence of the
 spatial part
 $\vec{X}(t)$ of the process  to  the It\^{o} diffusion process
 with infinitesimal increments :
 $${d\vec{X}(t) = [\vec{w}(\vec{x},t) + {1\over \beta}\vec{K}]dt
 +  \sqrt{2D} d\vec{W}(t)\enspace .}\eqno (15)$$

Consequently, the forward drift  of the process would read
$\vec{b}(\vec{x},t)= \vec{w}(\vec{x},t) + {1\over \beta }
\vec{K}(\vec{x})$. Notice that the
 $\beta ^{-1}\vec{K}$ contribution  can be safely ignored if
 we are interested in the  dominant driving motion.

Throughout the paper we are  interested in  Markovian
 diffusion processes, which propagate respectively the
 phase-space or
 configuration space probability densities (weak solutions
 of stochastic differential equations are thus involved).
 In the configuration space
 variant corresponding to  Eqs. (13), (15),
we deal with  a   stochastic  process whose probability
density $\rho (\vec{x},t)$
evolves according to  the standard Fokker-Planck equation
 $${\partial _t \rho =
 D\triangle  \rho - \vec{\nabla }\cdot (\vec{b}\rho )}
 \eqno (16)$$
which is supplemented by the momentum conservation law (in the mean)
 of the form (10) for $\vec{v}=\vec{b} - D{{\vec{\nabla }\rho }
 \over {\rho }}$.
If we compare Eq. (15) with  Eq. (7), we realize that  the
transformation  of drifts has been executed. Under suitable
restrictions, we can relate probability  measures
corresponding to those equations by means of the
Cameron-Martin-Girsanov theory    of measure transformations.
The Radon-Nikodym derivative of measures is here involved and
for suitable  forward drifts that
are gradient fields   it yields, \cite{blanch}, the most  general
 form of an auxiliary potential $\Omega (\vec{x},t)$
that is  allowed to appear in Eq. (10):
$${\Omega (\vec{x},t) = 2D[ \partial _t\phi + {1\over 2}
({\vec{b}^2\over {2D}} + \vec{\nabla }\cdot \vec{b})]\enspace .}
\eqno (17)$$
Here $\vec{b}(\vec{x},t) = 2D \vec{\nabla } \phi (\vec{x},t)$.

Eqs.  (17) and  (11) are  trivial identities,
if we take for granted that
all drifts are known from the beginning, like in case of typical
Smoluchowski diffusions where the external force $\vec{F}$ is
a priori postulated.
We can proceed otherwise and,
 on the contrary, one  can  depart
from a  suitably chosen    space-time dependent function
$\Omega (\vec{x},t)$.
Then Eq. (16) should be considered as  a nonlinear
(Riccatti-type) equation  which is to  be solved with
respect to the   drift field
potential    $\Phi (\vec{x},t)$. Such solution, when inserted to the
Fokker-Planck equation (16) would ultimately yield an evolution
of an initial probability density $\rho (\vec{x},0)$.

From this point of view, while developing the formalism, one should
decide what is a quantity of a \it primary \rm physical interest:
the field  of drifts $\vec{b}(\vec{x},t)$ or the potential
$\Omega (\vec{x},t)$.
They are not independent quantities, and enter the discussion
as entangled objects.   Mathematical features of the formalism
appear to depend crucially on the properties (like continuity,
local and global boundedness, Rellich class) of the potential
$\Omega $, see e.g. \cite{blanch}.

If we decide that  the momentum conservation law is governed by
a bounded from below, continuous function $\Omega (\vec{x},t)$,
cf. \cite{blanch}), then it  seems worthwhile to mention a close
 connection of the considered framework with  the general theory
  of  small random perturbations  of the
  classical Hamilton-Jacobi dynamics, \cite{zambrini}.
An   assumption that the forward  drift is  defined in terms of
a gradient of a suitable   function allows to rewrite the formula
(17) in the form clearly reminiscent of the
Hamilton-Jacobi equation (we set $\Phi = 2D\phi $ in Eq. (17)
and take $\Phi (\vec{x},0)$ as the initial data for the $t\geq 0$
evolution):
$${\Omega = \partial _t \Phi + {1\over 2}|\vec{\nabla }
\Phi |^2 +
D\triangle \Phi \enspace .}\eqno (18)$$
An associated function (known as the so-called backward drift of
a Markovian diffusion process) $\vec{b}_* = \vec{b} -2\vec{u}$,
cf. Eq. (11),  if denoted $\vec{b}_* = \vec{\nabla }\Phi _*$
is known  to yield
another   modified Hamilton-Jacobi equation, \cite{blanch,zambrini}:
 $${\Omega = \partial _t \Phi _*
+ {1\over 2} |\vec{\nabla } \Phi _*|^2 - D\triangle \Phi _*}
\eqno (19)$$
to be solved with given terminal data $\Phi _*(\vec{x},T)$ for times
$0\leq t\leq T$.

Eq. (19) is identifiable as the so-called Hamilton-Jacobi-Bellmann
programming equation in the optimal control of  stochastic diffusion
processes, \cite{zambrini}. A related  issue of viscosity
solutions of the  standard Hamilton-Jacobi equation has been
extensively studied in the literature as the
$D\downarrow  0$ limit of solutions of the modified (e.g. Bellmann)
equation. A direct connection (via the logarithmic Hopf-Cole
transformation) of Eq. (19) with the forced Burgers equation and the
generalised heat equation (hence, with the related Feynman-Kac
potentials, semigroups and kernels) is well known,
\cite{blanch,zambrini,muratore}.

There  is however  more interesting to observe that a gradient
field ansatz for the diffusion current velocity
($\vec{v}=\vec{\nabla }S$):
$${\partial _t\rho = - \vec{\nabla }\cdot [(\vec{\nabla } S)\rho ]}
\eqno (20)$$
allows to transform the momentum conservation law (10) of a
Markovian diffusion process to  the universal
Hamilton-Jacobi form:
$${\Omega = \partial _tS + {1\over 2} |\vec{\nabla }S|^2  + Q }
\eqno (21)$$
where $Q(\vec{x},t)$ was defined before in Eq. (12). By
applying the  gradient operation to Eq. (20) we recover (10).
Notice that Eq. (20) is sensitive to  any additive
(constant or time-dependent) modification of the potential $\Omega $.
In the above,  the contribution due to $Q$
is a direct consequence of  an initial probability measure choice
for the diffusion    process,
 while $\Omega $ via Eq. (17) alone does account for an appropriate
forward drift of the process.

The  derivation of a hierarchy of local
conservation laws (moment equations)  for the Kramers
equation can
be patterned after the standard procedure for the Boltzmann equation,
\cite{dorfman,wil,mazur}.
Those laws do not form a closed system and
additional specifications (like the familiar thermodynamical
equation of state) are needed to that end.
In case of the isothermal Brownian motion, when considered
in the large friction regime (e.g. Smoluchowski diffusion
approximation), the Fokker-Planck equation must be
supplemented by one
conservation law  \it only \rm to yield a closed system. Such
system  uniquely determines the stochastic process.

That happens under a definite choice of  external forces,
and hence Smoluchowski  drifts.
If the drifts are  not a priori  specified, then the only
freedom left in the momentum conservation law amounts to
the   choice of a concrete functional form for  the potential
 $\Omega (\vec{x},t)$.  In the theory of Brownian motion this
 particular decision making  replaces the standard
equation of state  constraint, suitable for the kinetic theory
description of gases and liquids.

In view of more sensitive dependence on the potential and hence
more detailed discrimination between distinct dynamics scenarios,
we adopt the Hamilton-Jacobi equation  (21) as  a  generic
substitute of the momentum conservation law (10).
Thus, we can consider a  closed system which is
composed of  the continuity
equation $\partial _t \rho =- \vec{\nabla }(\vec{v}\rho )$
(this,  in view of $\vec{v} =
\vec {b} - D{{\nabla \rho }\over {\rho }}$, is equivalent to
the Fokker-Planck equation (16))
 and  the Hamilton-Jacobi  equation
(21), plus suitable initial (and/or boundary) data.
Alternatively, we can supplement the Fokker-Planck equation  (16)
by the nonlinear Riccatti-type equation (17) to form a closed
system of partial differential equations, provided the
functional form of $\Omega $ is pre-selected.
In contrast to the pair (16), (21) where $\rho (\vec{x},t)$
enters an entangled relationship, the pair  (16), (17) is not
entangled.

We need to stress that it is the  closed system of Eqs.
(20)  and (21) which directly  refers to physically motivated
local conservations laws (moment equations) associated with
the Brownian motion, \cite{geilikman,mazur,blanch},
and to the respective diffusion currents.
The  underlying Markovian diffusion process is then specified
\it   uniquely \rm (that would \it not \rm be the case if
(10) has been used  instead of (21)).
However, Eqs. (20) and (21) form a coupled nonlinear system,
whose analytic solutions  are not readily accessible.
It is therefore useful to know  that  a  \it
linearisation \rm of this formidable
 nonlinear problem is provided by a time-adjoint pair of
generalised diffusion equations (cf. the Appendix) in the framework
 of the Schr\"{o}dinger boundary data problem.   The standard
heat equation appears as a very special  case in this
 formalism.

\section{The third Newton law in the mean}

\subsection{Free Brownian motion in terms of local conservation laws}

 The local conservation law  (10)  acquires a direct
 physical meaning (the
rate of change of momentum
carried by a locally co-moving with the flow volume, \cite{blanch}),
only if averaged with respect to $\rho (\vec{x},t)$ over
a   simply connected spatial area.   Namely, if $V$ stands for
a volume enclosed by a two-dimensional outward oriented
surface $\partial V$, we define a co-moving
volume on small time scales, by deforming the boundary surface in
accordance with the local current velocity  field values.
Namely, we consider  at time $t$ the  displacement of the
boundary surface $\partial V(t)$ defined  as follows:
$\vec{x}\in \partial V \rightarrow \vec{x} +
\vec{v}(\vec{x},t)\triangle t$ for all $\vec{x}\in \partial V$.
Up to the first order in $\triangle t$  this guarantees the
conservation of mass (probability measure)  contained
in $V$ at time $t$ i. e. $\int_{V(t+\triangle t)}\rho
(\vec{x},t+\triangle t)d^3x  - \int_{V(t)}\rho (\vec{x},t)d^3x
\sim 0$.

The  corresponding (to the leading order in $\triangle t$)
quantitative  momentum   rate-of-change measure reads,
cf. \cite{blanch},
$\int_V \rho \vec{\nabla }(\Omega - Q)d^3x$.
In view of  $\partial _iQ=
{1\over \rho }\sum_{j} \partial _jP_{ij}$,  where the stress
tensor $P_{ij}=D^2 \rho \partial _i \partial _j\,  ln\, \rho $ is
determined up to an additive  time-dependent or constant term,
the standard  divergence theorem allows to isolate an explicit
surface (due to stresses or of the pressure-type)
contribution. Namely, there holds
  $-\int_V\rho \partial _i Qd^3x=
-\int_{\partial V}\sum_{j} P_{ij}d\vec{\sigma }_j$,
with $d\vec{\sigma }$ being an infinitesimal area element
of $\partial V$ in $R^3$.

 For a particular case of the  free  Brownian expansion:
$${\rho _0(\vec{x})={1\over {(\pi \alpha ^2)^{3/2}}} exp
(-{x^2\over \alpha ^2}) \rightarrow   \rho(\vec{x},t)=
{1\over {[4\pi D(t+t_0)]^{3/2}}}\,
exp[-{{\vec{x}^2}\over {4D(t+t_0)}}]}\eqno (22)$$
where $\alpha ^2=4Dt_0$ and  $t_0>0, t\geq 0$,  we would
have
$${P_{ij}(\vec{x},t)= P(\vec{x},t)\delta _{ij}= -{{D}
\over {2(t+t_0)}}
\rho (\vec{x},t) \delta _{ij}}\eqno (23)$$
for all
$\vec{x}\in R^3$ and $t \geq 0$. Here $\delta _{ij}$ stands for the
Kronecker  symbol.   Then, $ -\int_V\rho \vec{\nabla }Q d^3x=-
\int_{\partial V}Pd\vec{\sigma }$, where
$${Q(\vec{x},t) =
{\vec{x}^2\over {8(t+t_0)^2}} - {{3D}\over {2(t+t_0)}}\enspace .}
\eqno (24)$$

The current velocity   $\vec{v}(\vec{x},t)=\vec{\nabla }S(\vec{x},t)=
{\vec{x}\over {2(t+t_0)}}$, apart from solving
$\partial _t\rho =-\vec{\nabla }\cdot (\vec{v}\rho )$
and $\partial _t\vec{v} + (\vec{v}\cdot \vec{\nabla })\vec{v} =
-\vec{\nabla } Q$  with $\rho _0$ and $\vec{v}_0=-\vec{u}_0$
standing for initial data,   is also linked to the Hamilton-Jacobi
equation
$${\partial _tS +{1\over 2}|\vec{\nabla }S|^2 + Q =0}\eqno (25)$$
whose solution is: $S(\vec{x},t)= {\vec{x}^2\over {4(t+t_0)}}
 + {3\over 2}D ln[4\pi D(t+t_0)]$.

Let us observe that the initial data $\vec{v}_0=-D\vec{\nabla }
ln\, \rho _0= -\vec{u}_0$ for the current
velocity field  indicate that we have totally  ignored
a crucial  \it preliminary \rm stage of the dynamics on the
$\beta ^{-1}$ time scale,  when the Brownian expansion of an
initially \it static \rm  ensemble has been ignited and so particles
have  been ultimately  set  in motion.

Notice also that our "osmotic expansion pressure" $P(\vec{x},t)$
is not positive definite, in contrast to
 the familiar kinetic theory (equation of state) expression
 for the pressure  $P(\vec{x})= \alpha \,
 \rho ^{\beta }(\vec{x}), \,  \alpha >0$ appropriate for gases.
 The admissibility of the negative sign of the "pressure"  function
 encodes the fact that the
Brownian evolving concentration of particles generically
decompresses (blows up),
instead of being compressed by the surrounding medium. The
compression (pressure upon the control volume coming from its
surrounding) is the standard feature in the kinetic theory of
gases, except for the
cavitation phenomenon in liquids and  the exotic  blow-up
conditions in  the concentrations of dense hot matter.

The loss  (in view of the "osmotic" migration) of momentum
 stored in a  control volume at a given time,
may be here interpreted
in terms of  an acceleration   $-\int_V\rho \vec{\nabla }Qd^3x$
induced by a \it fictituous \rm "attractive force".
By invoking an  explicit Hamilton-Jacobi connection (21),
we may attribute
to a diffusing Brownian  ensemble floating through a
(locally co-moving) control volume $V$,
 the mean kinetic energy   per unit of mass
$\int_V\rho {1\over 2}\vec{v}^2 d^3x$.
We can also evaluate the mean total kinetic  energy per unit of mass
obtained after extending integrations from $V$ to $R^3$.

For the
considered example, in view of $<\vec{x}^2>=6D(t+t_0)$,  we have
$\int_{R^3}\rho {1\over 2}\vec{v}^2 d^3x= {{3D}\over {4(t+t_0)}}$.
Notice that the mean energy
$\int_V\rho ({1\over 2}\vec{v}^2+Q) d^3x$  needs not to be  positive.
Indeed, this expression identically vanishes after extending
integrations from $V$ to $R^3$.
On the other hand the kinetic contribution, initially equal
$\int_{R^3}{1\over 2} \rho v^2 d^3x = 3D/\alpha ^2$ and
evidently coming from nowhere, continually
diminishes and is bound to disappear in the asymptotic
$t\rightarrow \infty $ limit, when  Brownian particles become
uniformly distributed in space.

\subsection{The third Newton law in the mean and the Brownian
recoil principle}

Normally, diffusion processes yielding a nontrivial matter
transport (diffusion currents) are observed
for a  non-uniform  concentration of colloidal particles.
We can devise a thought (numerical) experiment that gives rise
to a corresponding transport in terms of an ensemble of
sample Brownian motion realisations on a
fixed finite  time interval,
 instead of considering a multitude of them (migrating swarm of
 Brownian particles)  simultaneously.
One  may surely  implant
particles at initial (random) space locations to mimic
a certain probability density and next release (individually
in each sample experiment) and  allow them
to execute their Brownian paths independently, in
a fixed  duration time interval.
In terms of such a particle ensemble, we can safely return back
to the previous colloidal diffusion  picture where migrating
Brownian particles are  also regarded as independent (and so are
their individual Brownian motions).

 Consequently, in both visualizations, after the relaxation
 time $\beta ^{-1}$, the diffusion current   is initiated
 "at the expense" of the bath.
A nonzero mean kinetic energy must have been initially
transferred (pumped) from the bath to the diffusing
(blowing-up, expanding) swarm of particles.  Only asymptotically
this energy is being returned back to the bath.

Recalling our previous discussion,  for a sufficiently fast diffusion
process, all that should correspond to a local cooling of the bath
and implement a tiny deviation from its thermal equilibrium
conditions in each single particle propagation (simulation) experiment.
Accordingly,
the tendency to regain the local thermal equilibrium
by the bath (reflecting an attraction of tagged particles
by the cooler areas) must result in induced local flows - they
can become identifiable only on the  ensemble average.
All that is to happen \it not \rm literally  on the $\beta ^{-1}$
relaxation time scale but on (still relatively small)
time scale $\triangle t$ of the diffusion process which is well
below the observational one.

To each executed sample path there corresponds
a  sample realisation of the random medium (pushed
slightly away from its thermal  equilibrium, in view of the
postulated  feedback mechanism).
Those random (sample) realisations of the bath
 should be ensemble-averaged as well to yield
  an "effective" bath in  thermal equlibrium,  which is
  however    no  longer in a statistical state of rest.
  The emergent  driving flows mimic, on the average,
 the "return to equilibrium"  of the bath in each  sample
 propagation experiment.
The thermal conditions are  maintained on the ensemble average,
so that the \it effective \rm process guiding  the
\it ensemble \rm dynamics can be  viewed as  isothermal.

 We recall   close links of this scenario
 with an idea of a random walk in a  random medium,
 \cite{havlin}.
 However, presently a disorder in the random medium is coupled to
 the randomness of the walk and is no longer of independent
 origin.

 Let us assume that
"an effort" (hence, an energy loss) of the random medium, on the
$\beta ^{-1}$ scale, to produce a local Brownian diffusion current
$\vec{v}(\vec{x},t_0)$  out of the  initially static ensemble
and thus to decompress (lower the blow-up tendency) an initial
non-uniform probability distribution, results in the \it effective
osmotic   reaction \rm of  the random medium.
 Whatever is being transported away (on the ensemble average)
 according to the Fick law, is
 assumed to  induce a compensating osmotic  counterflow in the
 \it effective thermal bath.  \rm  That is the Brownian recoil
 effect of Ref. \cite{vigier}.

Now, the particle swarm   propagation scenario becomes
entirely different from the standard  one,  (10), (20).
First of all, the nonvanishing forward drift   $\vec{b}=\vec{u}$
is  generated as a dynamical (effective, statistical here !)
response of the bath to the enforced by the bath  particle
transport with the local
velocity $\vec{v}= -\vec{u}$.
Second, we need to account for a parellel  inversion of
the pressure effects (compression $+\vec{\nabla }Q$ should
replace the decompression $-\vec{\nabla }Q$) in the respective local
momentum conservation law.

Those features can be secured    through an explicit  realization
 of  the action-reaction principle
("the Brownian   recoil" effect, cf. Ref. \cite{vigier}),
 which we promote to the status of the \it  third Newton law
 in the mean. \rm

On the level of Eq. (10), once averaged over a finite
volume, we interpret the momentum per unit of mass rate-of-change
$\int_V \rho \vec{\nabla }(\Omega - Q)d^3x$   which occurs
exclusively due to the Brownian expansion, to generate a
counterbalancing rate-of-change tendency in the random medium.
To account for the emerging forward drift and an obvious
modification  of the subsequent dynamics of an ensemble of
(tagged) particles, we  re-define Eq. (10) by setting
  $- \int_V \rho \vec{\nabla }(\Omega - Q)d^3x$ in its
  right-hand-side instead of  $+ \int_V \rho \vec{\nabla }
  (\Omega - Q)d^3x$ .   That amounts to   an
  instantaneous  realisation of
  the third Newton law in the mean (action-reaction principle).
Hence, the momentum conservation law for the process \it
with a recoil \rm
(the reaction term replaces the decompressive "action" term)
would read:
$${\partial _t\vec{v} + (\vec{v}\cdot \vec{\nabla })\vec{v} =
\vec{\nabla } (Q- \Omega )}\eqno (26) $$
implying that
$${\partial _t S + {1\over 2}
|\vec{\nabla }S|^2 - Q= -\Omega }\eqno (27)$$
stands for the corresponding Hamilton-Jacobi equation, cf.
\cite{zambrini,misawa}, instead of Eq. (21).
A suitable adjustment (re-setting) of the initial data is
here necessary, which we shall explain in below
(cf. Section 3.3).

The major idea of the Brownian motion with  a recoil is thus given.

In the coarse-grained picture of motion we shall deal with a
sequence    of repeatable scenarios realised
on the  Smoluchowski process  time scale $\triangle t$:
the Brownian swarm expansion build-up is accompanied by the parallel
counterflow build-up, which in turn modifies the subsequent
stage of the Brownian swarm migration  (being interpreted to modify
the forward drift of the process) and the corresponding
built-up anew counterflow.

Although the new closed system of partial differential equations
(20) and (27) is  very different from the previous one (20), (21),
nonetheless it describes Markovian diffusion-type  processes
again, \cite{nel,blanch,carlen}.
The link is particularly obvious if we observe that the
 new Hamilton-Jacobi equation (27) can be formally
rewritten in the previous form (21) by introducing:
$${\Omega _r= \partial _tS + {1\over 2} |\vec{\nabla }S|^2 + Q }$$
$${\Omega _r= 2Q - \Omega }\eqno (28)$$
where $\Omega $ represents the previously defined potential function
of any Smoluchowski (or more general) diffusion process, (11).
It  is $\Omega _r$  which via Eq. (17) would determine forward drifts
of the Markovian diffusion process with a recoil. They must obey the
Cameron-Martin-Girsanov identity
$${\Omega _r = 2Q- \Omega =
2D[ \partial _t\phi + {1\over 2}
({\vec{b}^2\over {2D}} + \vec{\nabla }\cdot \vec{b})]\enspace .}
\eqno (29)$$

Our system of equations (20), (27) is badly nonlinear and coupled,
but its linearisation
can be immediately given in terms of an adjoint pair of
Schr\"{o}dinger equations with  a potential $\Omega $,
\cite{nel,zambrini}.
Indeed,
$${i\partial _t \psi = - D\triangle \psi + {\Omega \over {2D}}\psi }
\eqno (30)$$
with a solution
$${\psi = \rho ^{1/2} exp(iS)}\eqno (31)$$
and its complex adjoint makes the job, if
we regard  $\rho $ together with $S$ to remain in conformity with
the previous notations of Eqs. (20), (27). The choice of
$\psi (\vec{x},0)$ gives rise to a solvable Cauchy problem.
This feature we shall exploit in below.
Notice that, in view of Eq. (30), for time-indepedent $\Omega $, the
total energy $\int_{R^3}({v^2\over 2} -Q + \Omega )\rho d^3x$
 of the diffusing ensemble is a  conserved quantity.

The problem (20), (27), (28) can be
reformulated  as the Schr\"{o}dinger boundary data problem (cf. the
Appendix), but the  resulting   generalised diffusion equations are
nonlinearly  coupled (by means of the potential $\Omega _r$ replacing
the Feynman-Kac potential $\Omega $.
Hence the previous linearisation needs to be
exploited anyway.  The general existence criterions for Markovian
diffusion   processes of that kind, were formulated in Ref.
\cite{carlen}, see also \cite{zambrini,blanch} and the Appendix.

\subsection{Brownian motion with a  recoil as an anomalous (enhanced)
 diffusion model}

For the clarity of discussion, we shall confine our considerations
 to one-dimensional problems.
In the absence of external forces, we may consider a solution of
equations   (in  space dimension one)  $\partial _t\rho =
-\nabla (v\rho )$ and $\partial _t v + (v\nabla )v = + \nabla Q$,
where an initial probability density $\rho _0(x)$ is chosen in
correspondence with the previous free Brownian motion example.
We denote $\alpha ^2= 4Dt_0$.
Then,
$${\rho (x,t)=
{\alpha \over {[\pi (\alpha ^4 + 4D^2t^2)]^{1/2}}}\,
exp[-{{x^2\alpha ^2}\over  {\alpha ^4 + D^2t^2}}]}\eqno (32)$$
and
$${b(x,t)= v(x,t) + u(x,t)= {{2D(\alpha ^2 - 2Dt)x} \over
{\alpha ^4 +  4D^2t^2}}}\eqno (33)$$
are the pertinent solutions.
Notice that $u(x,0)= -{{2Dx}\over \alpha ^2}=b(x,0)$  amounts to
$v(x,0)=0$,  while in the previous free Brownian case the initial
current velocity was equal to $-D\nabla ln\, \rho _0$.
This re-adjustment of the initial data can be interpreted in
terms of the counterbalancing (recoil) phenomenon:
the  would-be initial Brownian
ensemble current velocity  $v_0=-u_0$ is here
completely saturated by the emerging forward  drift
$b_0=u_0$, see e.g. also \cite{vigier}.
 This   implies $u(x,t)= - {{2D\alpha ^2x}\over
{\alpha ^4 + 4D^2t^2}}$   and  $v(x,t)= {{4D^2xt}\over
{\alpha ^4 + 4D^2t^2}}$. Notice that $\nabla Q=  {x\alpha ^4
\over {4(\alpha ^4 +  D^2t^2)}}$, to be compared with the
respective expression $-\vec{\nabla }Q= - {{D^2x}
\over {\alpha ^2 + 4Dt}}$ in the previous section.
Presently, we deal  with a fictituous "repulsive" force,
which corresponds to the compression (pressure upon) of
the Brownian ensemble due to the counter-reaction of the
surrounding  medium.

We can write things more explicitly. Namely, now:
$${Q(x,t)= {{2D^2\alpha ^2}\over {\alpha ^4 + 4D^2t^2}}
({{\alpha ^2 x^2}\over {\alpha ^4+ 4D^2t^2}} - 1)}\eqno (34)$$
and the corresponding pressure term ($\nabla Q=
{1\over \rho }\nabla P$) reads
$${P(x,t) = - {{2D^2\alpha ^2}\over {\alpha ^4 + 4D^2t^2}}
\rho (x,t)}
\eqno (35)$$
giving a positive contribution $+\nabla Q$ to the local
conservation law (26).

The    related Hamilton-Jacobi equation
$${\partial _tS + {1\over 2} |\nabla S|^2 = + Q}\eqno (36)$$
is solved by
$${S(x,t) = {{2D^2x^2t}\over {\alpha ^4 + 4D^2t^2}} - D\,
arctan\, (-{{2Dt}\over \alpha ^2})}\eqno (37)$$

With  the above form of $Q(x,t)$ one can readily check that
equations (28) are identically satisfied, and
that the Cameron-Martin-Girsanov constraint euqation for the
forward drift of the Markovian diffusion process with a recoil
is automatically valid for $\phi  ={1\over 2}ln\, \rho  + S$:
$${2Q = 2D[\partial _t\phi +{1\over 2}({b^2 \over {2D}} +
\nabla \cdot b)]}\eqno (38)$$
cf. the general identity (29).

In anology with our free Brownian motion discussion, let us observe
that  presently
$${<x^2> = {\alpha ^2\over 2} + {{2D^2t^2}\over \alpha ^2}\enspace .}
\eqno (39) $$

It is easy to demonstrate (use a linearisation (30), (31) of the
problem) that the quadratic dependence on time
persists for arbitrarily  shaped  initial choices of the
probability distribution $\rho _0(x)>0$.
That signalizes an  anomalous behaviour (enhanced diffusion)
of the pertinent Markovian  process when $\Omega =0$
i. e. $\Omega _r=2Q$.

We can evaluate the kinetic energy contribution
$${\int_{R} \rho {v^2\over 2}dx = {{4D^4t^2}\over
{\alpha ^2(\alpha ^4 + 4D^2t^2)}}}\eqno (40)$$
which in contrast to the Brownian case shows up a continual growth
up to the terminal (asymptotic) value ${D^2\over \alpha ^2}$.
This value was in turn an initial kinetic  contribution
in  the previous Brownian example.
In contrast to that case, the total energy integral is now finite
(finite energy diffusions of Ref. \cite{carlen}) and reads
$${\int_R({1\over 2}v^2 - Q)\rho dx = {D^2\over \alpha ^2}}
\eqno (41)$$
for all times (it  is a conservation law).
The asymptotic value of the current velocity
$v\sim {x\over t}$  is twice larger than   this appropriate for the
Brownian motion, $v\sim {x\over {2t}}$.

\subsection{Response to an external force: non-dispersive
diffusion-type processes}

Let us regard $\vec{F}(\vec{x})=-\vec{\nabla }V(\vec{x})$  as
an external force field, whose  effects on the dynamics of
Brownian particles is encoded in the Smoluchowski diffusion
process equations (7), (9) and next (9)-(11). Those in turn
can be motivated by invoking the Kramers equation (14) and its
Smoluchowski diffusion  approximation (15).

As emphasized before, on the level of local conservation laws,
in the diffusion approximation, the microscopic force $\vec{F}$
is represented by the (Feynman-Kac)  potential $\Omega $,
defined through the Girsanov formula (11).

Let us adopt the third Newton law in the mean and the related
Brownian  recoil   strategy (26)-(29) to this case.
Evidently, the potential $\Omega $ will explicitly appear
in the  linearization (30) of the  problem.
On the other hand, it is
 the  potential $\Omega _r=2Q-\Omega $ which via Eq. (29)
 determines  forward drifts  appropriate for the diffusion
  process with a recoil.
 In view of an inherent nonlinearity of the problem,
 one should not expect that the emergent drifts  would allow for
 a simple decomposition  met in Eqs. (14), (15).

Our further discussion will be carried out in one space dimension
and will focus on quadratic potentials.

For a parabolic (harmonic oscillator) potential $V(x)=
{1\over 2}m \omega ^2 x^2$ defining the acceleration
$K(x)= -\omega ^2x$, the corresponding
Feynman-Kac potential (11)
reads $\Omega (x)={1\over 2}\gamma ^2x^2   - D\gamma , \, \gamma =
{\omega ^2\over \beta }$.
It is useful to mention that the choice of the repulsive potential
$V(x)= + {1\over 2}m \omega ^2x^2$ would yield an innocent-looking
modification  by a constant in the function (17):
$\Omega (x)={1\over 2}\gamma ^2x^2 + D\gamma $.
That demonstrates an extraodrinary sensitivity  of the
Riccatti-type equations  (11) and (17) on the choice
of $\Omega $.

In fact, a suitable   additive modification of
${1\over 2}\gamma ^2x^2$ by a constant,
allows to generate (by directly solving the Riccatti-type
equation (11)) the whole family  of forward drifts pertaining to
\it inequivalent \rm stationary diffusion processes,
cf. \cite{blanch}.
Nonetheless, all of them
correspond   to the same $\nabla \Omega =+ \gamma ^2x$ generic
contribution to the local momentum conservation law (10).
Clearly,  the   law (10) does not provide a sufficiently fine
discrimination between admissible stochastic motion
scenarios, unless we know the Smoluchowski force and its
potential from the beginning.
It is only the Hamilton-Jacobi equation level,
where the closed system of partial differential equations
(respectively (20), (21) or (20), (27)) determines the
process uniquely.

It is clear that stationary processes \it are the same \rm
both in case of  the  standard Brownian motion  and the
Brownian motion with  a recoil. The respective
propagation scenarios substantially  differ in the
non-stationary case only.

To exemplify the above statement (we have discussed before
 the  $\Omega =0$ case), let us consider an explicit solution
of Eqs. (20) and (28) in case of $\Omega (x)=
{1\over 2}\gamma ^2x^2   - D\gamma $.
By means of the linearisation (30), (31), this can be easily
accomplished, cf. \cite{misawa}.
We shall utilse exactly the same inital probability density
$\rho _0(x)$ as before.
We have:
$$ {\rho (x,t) = [{{\gamma \alpha ^2}\over
{\pi (sin^2(\sqrt{\gamma }2Dt) +
\gamma \alpha ^4 cos^2(\sqrt{\gamma }2Dt))}}]^{1/2}  \cdot }$$
$${\cdot exp[-{{\gamma \alpha ^2 x^2}\over
{sin^2(\sqrt{\gamma }2Dt) +
\gamma \alpha ^4 cos^2(\sqrt{\gamma }2Dt)}}]}\eqno (41)$$
and
$${S(x,t)= - {{D(\gamma )^{3/2}\alpha ^4 x^2  tg\,
(\sqrt{\gamma }2Dt)}
\over {(tan^2(\sqrt{\gamma }2Dt + \gamma \alpha ^4)
 sin(\sqrt{\gamma }2Dt)}} + {{D\sqrt{\gamma } x^2}\over
 {tan(\sqrt{\gamma }2Dt)}}+}$$
 $${ + Darctan \,
 [-{{tan(\sqrt{\gamma }2Dt)}\over {\sqrt{\gamma }\alpha ^2}}] -
 D\gamma t\enspace .}\eqno (42)$$

The forward drift of the corresponding diffusion-type process reads:
$${b(x,t) =    {{(1 - \gamma \alpha ^2)\sqrt{\gamma }
sin(\sqrt{\gamma }4Dt)  - 2\gamma \alpha ^2}\over
{ sin^2(\sqrt{\gamma }2Dt) +
\gamma \alpha ^4 cos^2(\sqrt{\gamma }2Dt)}} D x}\eqno (43)$$
and $X(t)=x$ holds true in terms of the random variable of
the process.
The additive decomposition of the drift, valid in Eqs (9), (15),
  is completely destroyed by the Brownian recoil scenario.
Notice that  $b(x,0)= -{{2Dx}\over \alpha ^2} = u(x,0)$,
while (cf. (9)) $b={F\over {m\beta }}= -\gamma x$ would hold true
for all times, in case of the standard Smoluchowski
diffusion process.

Because  of the harmonic attraction  and suitable initial
probability measure choice, we have here
wiped out all  previously  discussed enhanced diffusion
features.
Now,  the dispersion  is attentuated and actually  the
non-dispersive diffusion-type  process is  realised:
$<x^2>$ does not spread at all despite of the intrinsically
stochastic  nature of the dynamics (finite-energy diffusions
of Ref. \cite{carlen}).

\vskip0.2cm
{\bf Acknowledgements:}\\
I would like to thank Professor Eric Carlen for discussion
about the scaling limits of the Boltzmann equation and  related
conservation laws. I am willing words of gratitude to Professor
Ana Bela Cruzeiro and Professor Jean-Claude Zambrini  for
a hospitality  at the University of Lisbon where part of this
work has been  completed.

\vskip0.5cm

{\bf Appendix: Reconstruction of a  Markovian diffusion  process
 from the input-output statistics data}
\vskip0.3cm
 There are many procedures to reproduce the intrinsic dynamics
 of a physical system from observable data, like e.g. the time
 series analysis.
 We shall outline  an algorithm allowing to reconstruct the  \it
 most likely \rm  microscopic  motion scenario  under an additional
 assumption that the sought for dynamics actually  \it is \rm
 a   Markovian diffusion process.   This reconstruction method
 is based on solving the so-called Schr\"{o}dinger boundary-data
 and interpolation problem, \cite{zambrini,blanch,muratore}.

 Given two strictly positive (usually on an open space-interval)
 boundary probability densities  $\rho _0(\vec{x}), \rho _T(\vec{x})$
 for a process with the time of duration $T\geq 0$.
One can  single out  a unique Markovian
diffusion process which is specified by solving  the Schr\"{o}dinger
boundary data problem:
$${m_T(A,B) = \int_A d^3x\int_B d^3y\,  m_T(\vec{x},\vec{y})}$$
$$\int d^3y\,  m_T(\vec{x},\vec{y}) = \rho _0(\vec{x}) $$
$$ \int d^3x\,  m_T(\vec{x},\vec{y})=\rho _T(y)$$
where  the joint probability distribution has a  density
$${m_T(\vec{x},\vec{y}) = u_0(\vec{x})\, k(x,0,y,T)\, v_T(\vec{y})}
0$$
and the two unknown  functions
$u_0(\vec{x}), \, v_T(\vec{y})$ come out as (unique) solutions, of
\it the same sign, \rm of the integral identities.
To this end, we need to have at our disposal a
continuous bounded  strictly positive (ways to relax this assumption
are known) function
$k(\vec{x},s,\vec{y},t),0\leq s<t\leq T$, which for our purposes
(an obvious
way to secure the Markov property) is chosen to be represented by
familiar Feynman-Kac integral kernels
 of contractive dynamical semigroup operators:
$${k(\vec{y},s,\vec{x},t)=\int exp[-\int_s^tc(
\vec{\omega }(\tau ),\tau)d\tau ]
d\mu ^{(\vec{y},s)}_{(\vec{x},t)}(\omega )}$$

 The pertinent   (interpolating) Markovian
process can be ultimately determined  by means of
positive solutions (it is desirable to have them bounded)
 of the adjoint pair of  generalised heat equations:
$${\partial _tu(\vec{x},t)=\nu \triangle u(\vec{x},t) -
c(\vec{x},t)u(\vec{x},t)}$$
$$\partial _tv(\vec{x},t)= -\nu \triangle v(\vec{x},t) +
c(\vec{x},t)v(\vec{x},t)\enspace .$$
Here, a function $c(\vec{x},t)$
is restricted only by the positivity and continuity demand
for the kernel.
In the above, $d\mu ^{(\vec{y},s)}_{(\vec{x},t)}(\omega)$ is the
conditional
Wiener measure over sample paths of the standard Brownian motion.

Solutions, upon suitable normalisation give rise to the
Markovian  diffusion process with the \it factorised \rm
probability density $\rho (\vec{x},t)=u(\vec{x},t)v(\vec{x},t)$
which, while evolving in time, interpolates between
the boundary density data $\rho (\vec{x},0)$ and 
$\rho (\vec{x},T)$. The interpolation admits an It\^{o} realisation
  with the respective forward and
backward drifts  defined as follows:
$${\vec{b}(\vec{x},t)=2\nu {{\nabla v(\vec{x},t)}
\over {v(\vec{x},t)}}}
$$
$$\vec{b}_*(\vec{x},t)= - 2\nu {{\nabla u(\vec{x},t)}
\over {u(\vec{x},t)}}$$
in the prescribed time interval $[0,T]$.\\
For the forward interpolation, the familiar Fokker-Planck
(second Kolmogorov) equation holds true:
$${\partial _t\rho (\vec{x},t) = \nu \triangle
\rho (\vec{x},t) - \nabla [\vec{b}(\vec{x},t)\rho (\vec{x},t)]}
$$
with $\rho (\vec{x},0)$ given, while for the backward
interpolation (starting from $\rho (\vec{x},T)$) we have:
$${\partial _t\rho(\vec{x},t) = - \nu \triangle \rho (\vec{x},t) -
\nabla [\vec{b}_*(\vec{x},t) \rho (\vec{x},t)]\enspace . }$$

The drifts are
gradient fields, $curl \, \vec{b}= 0$. As a consequence,
those  that are allowed by any  prescribed choice of   the function
$c(\vec{x},t)$  \it must \rm fulfill the compatibility
condition
$${c(\vec{x},t) = \partial _t \Phi \, +\,
{1\over 2} ({b^2
\over {2\nu }}+ \nabla b)}$$
which establishes the Girsanov-type  connection of
the forward drift
$\vec{b}(\vec{x},t)=2\nu \nabla \Phi (\vec{x},t)$ with the
Feynman-Kac potential  $c(\vec{x},t)$.
In the considered Schr\"{o}dinger's interpolation
framework, the forward and backward
drift fields  are  connected   by the identity
$\vec{b}_*= \vec{b} - 2\nu \nabla ln \rho $.

For  Markovian diffusion processes the
notion of the \it backward \rm transition probability density
 $p_*(\vec{y},s,\vec{x},t)$ can be consistently introduced on 
each finite  time interval, say $0\leq s<t\leq T$:
 $${\rho (\vec{x},t) p_*(\vec{y},s,\vec{x},t)=
 p(\vec{y},s,\vec{x},t) \rho (\vec{y},s)} $$
so that $\int \rho (\vec{y},s)p(\vec{y},s,\vec{x},t)d^3y=
\rho (\vec{x},t)$
and $\rho (\vec{y},s)=\int p_*(\vec{y},s,\vec{x},t)
\rho (\vec{x},t)d^3x$.  

The transport (density evolution) equations  refer to
processes running
in opposite  directions  in a fixed, common for both
time-duration period.
The  forward one executes an interpolation from the Borel set $A$
to $B$, while the  backward one executes  an interpolation from
$B$ to $A$.

The knowledge of the Feynman-Kac kernel  implies that the
transition probability density of the forward  process reads:
$${p(\vec{y},s,\vec{x},t)=k(\vec{y},s,\vec{x},t)
{{v(\vec{x},t)}\over {v(\vec{y},s)}}\enspace .}$$
while the corresponding
 transition
probability density  of the backward process has the form:
$${p_*(\vec{y},s,\vec{x},t) = k(\vec{y},s,\vec{x},t)
{{u(\vec{y},s)}\over {u(\vec{x},t)}}\enspace .}$$
Obviously in the time interval $0\leq s<t\leq T$
there holds:
$${u(\vec{x},t)=\int u_0(\vec{y}) k(\vec{y},s,\vec{x},t) d^3y}
$$
$$v(\vec{y},s)=\int k(\vec{y},s,\vec{x},T) v_T(\vec{x})d^3x
\enspace .$$
Consequently, we have fully determined the underlying  (Markovian)
random motions, forward and backward, respectively.
All that accounts for perturbations of (and conditioning upon)
the Wiener noise.\\

\vskip0.2cm
{\bf Remark 6:}
Various partial differential equations associated with Markovian
diffusion processes are known \it not \rm to be invariant under
time reversal (hence being dissipative and linked to irreversible
physical phenomena).
However,  the correspoding processes admit a \it statistical
inversion.   \rm
Let us consider a process running in a finite time interval,
say $[0,T]$. We may consistently define a process
running backward in time in this interval and
reproducing the most likely (statistical) past of the process,
given the present probability measure data.
See e.g. \cite{nel,vigier,blanch,zambrini} and \cite{math}.
In fact, cf. \cite{shreve} p. 255: " any probabilistic
treatement of the heat equation involves a time-reversal".
This feature is
explicitly  utilized in the analysis  of the above outlined
Schr\"{o}dinger  boundary-data and interpolation problem,
\cite{zambrini,blanch}.

\hspace*{1cm}

\end{document}